\documentclass[fleqn,10pt]{wlscirep}
\usepackage[utf8]{inputenc}
\usepackage[T1]{fontenc}
\usepackage[framed,numbered,autolinebreaks,useliterate]{mcode}

\title{Measuring Power with a Saturated Photodiode}

\author[1,*]{Shiekh Zia Uddin}
\affil[1]{Research Laboratory of Electronics, Massachusetts Institute of Technology, Cambridge, MA 02139, USA}
\affil[*]{suddin@mit.edu}

\begin{abstract}
Accurate measurement of optical power is pivotal in many applications and scientific research. However, traditional power meters are unable to measure power levels beyond a certain saturation point, limiting their usefulness in high-power applications. In this technical note, I discuss how optical power can be measured using a saturated photodiode. I demonstrate that by monitoring both the dc photocurrent and ac noise, it is possible to accurately measure power levels beyond its saturation point. 
\end{abstract}
\begin{document}

\flushbottom
\maketitle
%
%
\thispagestyle{empty}


\section*{Introduction}
Optical power measurement is a critical aspect of many applications. It is the conventional wisdom that a saturated photodiode can not be used to measure power. Saturation power of photodiodes can be pushed to higher levels by applying a reverse bias voltage, however there is a limit to the amount of bias voltage due to the reverse breakdown which can be catastrophic to the diode. This limitation can be problematic in high-power applications, where it is important to be able to accurately measure power levels at high speed. In this technical note, I discuss a method for measuring optical power using a saturated photodiode. I demonstrate that the photocurrent noise decreases with power beyond saturation which can be used to accurately measure power at levels beyond its saturation point. This information might be useful in photon noise measurements.

\section*{Background}
If a photoevent generated at $t = 0$ produces an electric pulse
$h(t)$, of area $e$, in the external circuit. A photoevent generated at time $t_1$ then produces a displaced
pulse, $h(t - t_1 )$. Dividing the time axis into incremental time intervals $\Delta t$ so that the probability $p$ that
a photoevent occurs within an interval is $P =\eta \Phi \Delta t$. The electric current $i$ at time $t$ is written as
\begin{equation}
  i(t)= \sum_l X_l h(t - l\Delta t),  
\end{equation}

where $X_l$ assumes the value $1$ with probability $p$, and $0$ with probability $1 - p$. The variables ${X_l}$
are independent. The mean value of $X_l$ is $E[X_l]=0\times(1 - p) + 1 \times p = p$. Its mean-square value is
$E[X_l^2] = 0^2 \times (1 - p) + 1^2 \times p = p$. The mean of the product $X_l X_k$ is $p^2$ if $l \neq k$, and $p$ if $l = k$.
The mean and mean-square values of $i(t)$ are now determined via
\begin{align}
    \overline{i}=E[i(t)] &=\sum_l p h(t - l\Delta t),  \\
    E[i^2(t)] &= \sum_l \sum_k E[X_l X_k] h(t - l\Delta t)h(t - k\Delta t) \\
    &=\sum \sum_{l\neq k} p^2 h(t - l\Delta t)h(t - k\Delta t) + \sum_l p h^2(t - l\Delta t)
\end{align}
Substituting $p = \eta \Phi \Delta t$, and taking the limit $\Delta t \rightarrow 0$ so that the summations become integrals, previous equations yield, respectively,
\begin{align}
    E[i(t)] &=\eta \Phi \int h(t)dt,  \\
    E[i^2(t)] &= \left( \eta \Phi \int h(t)dt \right)^2 + \eta \Phi \int h^2(t)dt
\end{align}
The limits of the integration is zero to infinity. It follows that
\begin{equation}
    \sigma_i^2=E[i^2]-E[i]^2= \eta \Phi \int h^2(t)dt
\end{equation}
Definition of the bandwidth $B$ as 
\begin{equation}
    B=\frac{1}{2e^2}\int_0^\infty  h^2(t)dt = \frac{\int_0^\infty  h^2(t)dt }{2 \left( \int_0^\infty  h(t)dt \right)^2},
\end{equation}
can be readily verified by noting that the Fourier transform of $h(t)$ is its transfer function $H(v)$. The area under $h(t)$ is simply $H(0) = e$. In accordance with Parseval's theorem, the area under $h^2(t)$ is equal to the area under the symmetric function
$|H(v)|^2$, so that 
\begin{equation}
    B=\int_0^\infty \left| \frac{H(v)}{H(0)} \right|^2dv
\end{equation}

The quantity $B$ is therefore the power-equivalent spectral width of the function $H(v)$
(i.e., the bandwidth of the device/circuit combination).
As an example, if $H(v) =1$ for $-V_c < v < V_c$ and $0$ elsewhere, we get $B =V_c$. Using this definition of bandwidth, we get back our familiar expression for the noise in photocurrent
\begin{equation}
    \sigma_i^2=2eE[i]B.
\end{equation}
So far this is a standard derivation of the shot noise \cite{saleh2019fundamentals}. Note that this expression hinges on the assumption that the area under $h(t)$ is simply $e$, basically one photoevent cause one electrons worth of charge to flow as current. However in saturation its definitely not the case and the photodiode response becomes a function of intensity. In the first order approximation, one can make the assumption that $h(t)=f(\Phi)g(t)$, where $f(\Phi)$ is a power dependent function that is $1$ at low power and decreases at high power and area under $g(t)$ is $e$. Then we get
\begin{align}
    E[i] &= \eta \Phi f(\Phi) \int g(t)dt \\
    \sigma_i^2 &=\eta \Phi f^2(\Phi) \int g^2(t)dt,
\end{align}
which gives us the key insight that the average value of current and noise power scales differently with incident optical power. If we take a ratio
\begin{equation}
    \frac{E[i]^2}{\sigma_i^2} = \frac{\eta^2 \Phi^2 f^2(\Phi) e^2 }{\eta \Phi f^2(\Phi) 2e^2B} \propto \Phi
\end{equation}
we find that the signal to noise ratio (SNR) in principle is proportional to incident power despite the nonlinearity in the response. Even if this proportionality does not hold exactly, with proper calibration therefore it should be possible to measure the power with a saturated photodiode by measuring both the average photocurrent and the photocurrent noise.
\section*{Results}
\begin{figure}[p]
\centering
\includegraphics[width=7in]{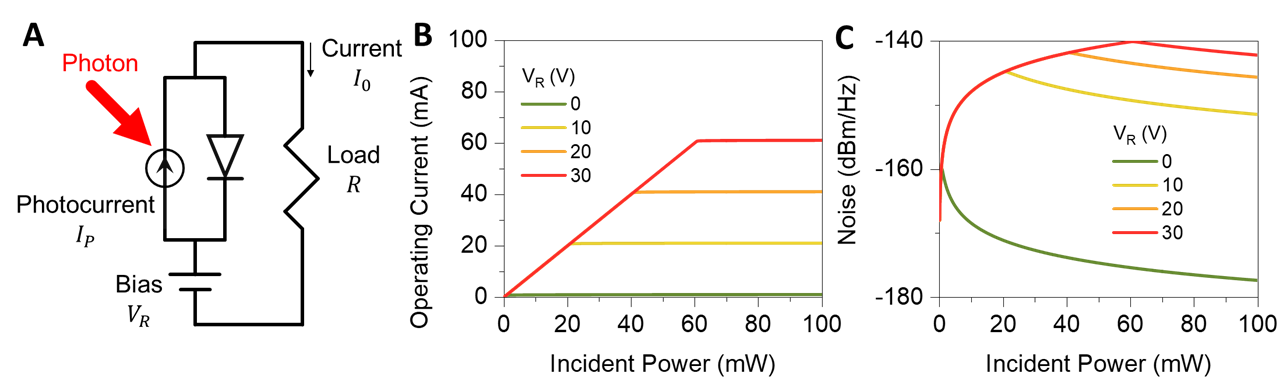}
\caption{Schematic diagram of a photodiode driving circuit and simulated operating current $I_0$.}
\label{F1}
\end{figure}

Figure \ref{F1}A shows a schematic diagram of a standard photodiode driving circuit. The photodiode can be modelled as a current source in parallel with a diode\cite{bhattacharya1997semiconductor}. A reverse bias voltage is applied to set the operating point, but there is a limit to how much voltage can be applied defined by junction breakdown\cite{neamen2003semiconductor}. Such a circuit can be simulated using conventional electrical circuit theory \cite{sedra2004microelectronic} (MATLAB code below), the resulting operating current with realistic circuit parameters is shown in Fig. \ref{F1}B. We can see that the photodiode saturates at some power and the saturation knee increases with reverse bias voltage. The highest measurable power is determined by the highest reverse bias voltage that can be applied across a junction. Below saturation the current is linear with incident power, which is expected.

\begin{figure}[p]
\centering
\includegraphics[width=3in]{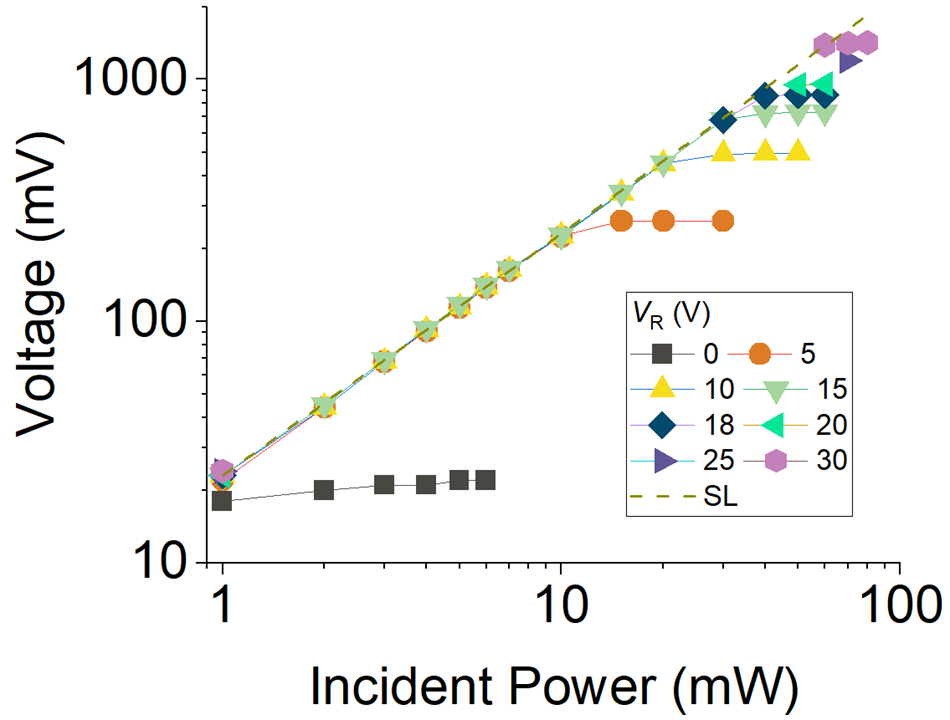}
\caption{Output voltage across the $25\Omega$ load resistance from a Thorlabs FDGA05 \cite{thorlabs} InGaAs photodiode at different reverse bias excited by a $1070$ nm CW laser. It has a $0.95$ A/W responsivity, $2.5$ ns rise time and $0.5$ mm active area diameter.}
\label{F2}
\end{figure}

Experimental data of a reverse biased photodiode is shown in Fig. \ref{F2} where reverse bias is seen to increase the saturation power. Before saturation the voltage is proportional to power. After saturation no measurement of power is possible, which is the current paradigm.

\begin{figure}[p]
\centering
\includegraphics[width=6in]{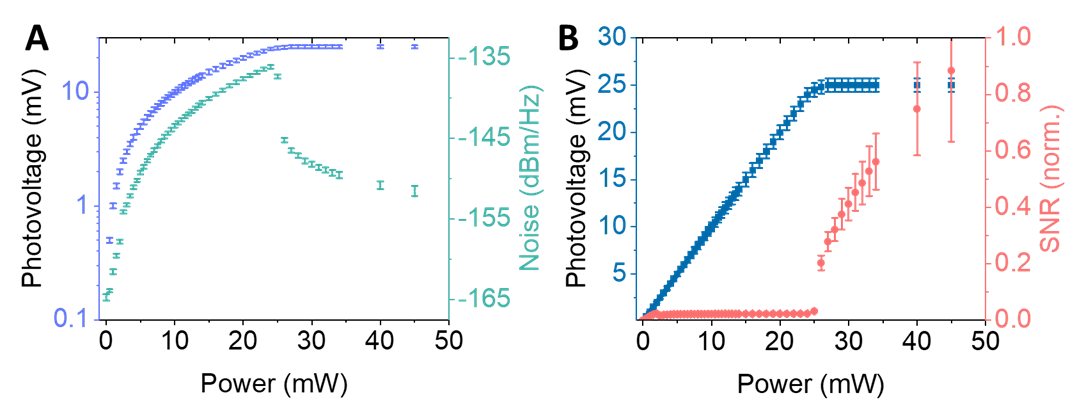}
\caption{(A) Output voltage and noise from a Excelitas C30641GH \cite{Excelitas} InGaAs photodiode at $30$ V reverse bias excited by a $1550$ nm femtosecond pulsed laser. It saturates around $25$ mW. (B) Even though photovoltage has saturated, the SNR shows response beyond the saturation power.}
\label{F3}
\end{figure}

We can now attempt to measure power beyond saturation. In Fig. \ref{F3}A we show the photovoltage and the photocurrent noise as a function of incident power. Photocurrent noise is measured around $1.5$ MHz with a spectrum analyzer ($10$ kHz resolution bandwidth, $1$ MHz span, with preamplifier on and no attenuation, electronic noise floor is $-165$ dBm/Hz). Below saturation photovoltage and noise increases simultaneously. As the photovoltage is saturated, the photocurrent noise suddenly decreases. This qualitatively follows our theoretical voltage and noise shown in Fig. \ref{F1}. In Fig. \ref{F3}B we show the photovoltage and SNR calculated from the experimental data. Beyond saturation SNR changes with incident power, which can be used to measure power after proper calibration. Such behaviour also holds at other noise frequencies as long as they are lower than the badngap of the photodiode and away from $1/f$ noise.

.

\section*{Discussion}

Even when a photodiode is saturated, the information about the photon flux intensity are not completely lost and in a way encoded in the photocurrent noise, which can be practically used to measure power at high speed after proper calibration.

\section*{Codes}

\begin{lstlisting}
%% Matlab code to solve for photocurrent and noise in a circuit
clc;clear all;close all;
P=linspace(0,100,1000)*1e-3; % Incident power in W
Responsivity=1;
Ip=P*Responsivity; % Expected photocurrent
R=500;
V=[linspace(-50,0,1e5),linspace(0,.7,1e5)];
VR=30; % Reverse Bias Voltage
Iop=zeros(size(P)); % Operating Photocurrent

for indx=1:length(P)
    I1=-Ip(indx)+0.1e-9*exp(V/.0259);
    I2=-(V+VR)/R;
    [~,pos]=min(abs(I1-I2));
    Iop(indx)=-I2(pos); 
end

figure(1);subplot(121), plot(P/1e-3,Iop); % Operating Current vs Power
f=Iop./P; %nonlinear response function
S=10*log10(2*1.6e-19*R*1*(P/1e-3).*(f.^2)); % Noise power vs optical power
subplot(122), plot(P/1e-3,S);
\end{lstlisting}

\bibliography{main}

\section*{Acknowledgements}
The author acknowledges Nicholas Rivera, Jamison Sloan, Yannick Salamin, chatGPT for their discussions. All equipment used in the experiments are properties of MIT.


\end{document}